\documentclass[preprint,superscriptaddress,showpacs]{revtex4}
\usepackage{epsfig} 
\usepackage{epstopdf}
\usepackage{hyperref} 
\usepackage{cleveref} 
\usepackage{graphicx} 
\usepackage{amssymb, amstext, amsmath} 
\usepackage{color}
\usepackage{verbatim}

\newcommand{\iu}{\mathrm{i}} 
\newcommand{\po}{\widetilde}
\newcommand{\NormConst} {\kappa}
\newcommand{\PME}{PTRE}
\newcommand{\intpic}{\widehat}            

\begin{document} 
\title{Coherent quantum transport in disordered systems: 
A unified polaron treatment of hopping and band-like transport}
\author{Chee Kong Lee}
\affiliation{Department of Chemistry, Massachusetts
Institute of Technology, Cambridge, Massachusetts 02139, USA}
\affiliation{Centre for
Quantum Technologies, National University of Singapore, 117543, Singapore} 

\author{Jeremy Moix} 
\affiliation{Department of Chemistry, Massachusetts
Institute of Technology, Cambridge, Massachusetts 02139, USA}
\author{Jianshu Cao} \email{jianshu@mit.edu}
\affiliation{Department of Chemistry, Massachusetts
Institute of Technology, Cambridge, Massachusetts 02139, USA}
\date{\today} 

\begin{abstract} 
Quantum transport in disordered systems 
is studied using a polaron-based master equation.
The polaron approach is capable of bridging the results from the coherent
band-like transport regime governed by the Redfield equation to incoherent
hopping transport in the classical regime.
A non-monotonic dependence of the diffusion coefficient is observed
both as a function of temperature and system-phonon coupling strength. 
In the band-like transport regime, the diffusion coefficient 
is shown to be linearly proportional to the system-phonon 
coupling strength, 
and vanishes at zero coupling due to Anderson localization.
In the opposite classical hopping regime, we correctly recover that the
dynamics are described by the 
Fermi's Golden Rule (FGR) and
establish that the scaling of the diffusion coefficient depends 
on the phonon bath relaxation time. 
In both the hopping and band-like transport regimes, it is demonstrated that
at low temperature the zero-point fluctuations of the bath lead to
non-zero transport rates, and hence a finite diffusion constant.
Application to rubrene and other organic semiconductor materials shows a 
good agreement with experimental mobility data.

\end{abstract}
\maketitle

\section{Introduction} 
Quantum transport in disordered systems governs a host of fundamental physical 
processes including the efficiency of light harvesting systems, 
organic photovoltaics, conducting polymers, and J-aggregate thin 
films~\cite{podzorov04, troisi06, sakanoue10,  %
singh09, dykstra08, bednarz03, moix11, Lee2007, %
Mayou2006, Coropceanu2007, Ortmann2009, Ciuchi2011, tao12a, cheng08}.
However, our theoretical understanding of these processes is still lacking
in many cases.
At the most basic level, one may describe the energy transport
as a quantum diffusion process occurring in a system that is 
influenced by both static disorder and thermal fluctuations.
Theories based on FGR, i.e.~the Marcus or F{\"o}rster 
rate expressions, are often used to study the transport properties of 
disordered organic systems,
but only in very few cases, such as the transport in 
organic crystals,
do the dynamics reside in a regime that is amenable to perturbative 
treatments~\cite{bednarz04}. 
More commonly, the coupling between the excitonic system and
the environment is neither large nor small so that these perturbative
treatments often yield qualitatively incorrect results.
This is particularly true in the case of biological light harvesting complexes
which are among the most efficient energy transporting systems
currently known.

Recently, the Haken-Strobl model~\cite{Moix2013a} and 
an approximate stochastic Schr{\"o}dinger equation~\cite{Zhong2014}  
have been used to study the energy transport processes in 
one-dimensional disordered systems.
However, the Haken-Strobl model represents a vast simplification of the 
true dynamics that is applicable only in the high temperature Markovian limit, 
while the approximate stochastic Schr{\"o}dinger equation is only 
valid in the weak system-phonon coupling regime and fails to correctly 
reproduce the classical hopping dynamics at high temperature.
Here we present a complete characterization of the transport properties 
over the entire range of phonon bath parameters through the development
of an efficient and accurate polaron-transformed Redfield master equation 
(\PME).
In contrast to many standard perturbative treatments, the \PME\  
allows one to treat systems that are strongly coupled to the
phonon baths.
Furthermore, the \PME\ is still very accurate even
if the system and bath are not strongly coupled provided that the bath 
relaxation time is sufficiently short~\cite{Lee2012, Chang2013}.
This approach allows us to explore many interesting features of the
dynamics that were previously inaccessible.

In the high temperature, incoherent regime, we recover the known scaling 
relations of hopping transport that are obtained from FGR. 
If the bath relaxation time is fast, then the diffusion constant, $D$, 
decreases with temperature as $T^{-1}$ 
as was found in the previous Haken-Strobl analysis.
However, as the bath relaxation time slows, the 
FGR reduces to the Marcus rates, and the temperature scaling of the
diffusion constant transitions to $D\propto T^{-1/2}$.
However, these relations hold only 
in the high-temperature/strong-coupling limit where the dynamics is incoherent.
As the temperature or system-bath coupling decreases, quantum
coherence begins to play a role and the FGR results 
quickly break down, leading to a significant underestimation of 
the true transport rate. 
In the sufficiently weak damping regime, the \PME\ results
reduce to those of band-like transport governed by 
the standard secular Redfield equation (SRE),
wherein the diffusion coefficient can be shown to increase linearly with the
system-phonon coupling strength. 
The SRE rates also demonstrate that transport occurs 
--even at zero temperature-- provided that the system-phonon coupling is finite, 
due to the dephasing interactions from the phonon bath.

This paper is organized as follows. We describe the \PME\ 
used to compute the diffusion coefficient in an infinite, disordered
one-dimensional chain in Sec.~\ref{section:theory}. In the following Sec.~\ref{section:results},  
the numerical results are presented and compared with the results
from standard SRE and FGR approaches
in the weak and strong coupling regimes, respectively.
The limiting results allow for the accurate determination of the respective 
scaling relations of the diffusion coefficient in each case. 
The \PME\ is used to study some common organic semiconductor materials in Sec.~\ref{section:applications}. 
Finally, we conclude with a summary of the results in Sec.~\ref{section:conclusion}.

\section{Theory}
\label{section:theory}
The total Hamiltonian in the open quantum system formalism is given by
\begin{equation}
   H_{\rm tot}=H_{\rm s}+H_{\rm b}+H_{\rm sb},
\end{equation} 
where the three terms represent Hamiltonians of the system, the phonon bath, and
system-bath coupling, respectively. The system is described by a tight binding,
Anderson Hamiltonian 
$H_{\rm s} = \sum_n \epsilon_{n} |n\rangle \langle n| + \sum_{m\neq n} J_{mn} |m\rangle \langle n | $, 
where $|n\rangle$ denotes the site basis and $J_{mn}$ is the electronic
coupling between site $m$ and site $n$. 
Here we only consider one dimensional systems with 
nearest-neighbor coupling such that 
$J_{mn} = J(\delta_{m,n+1} +\delta_{m+1,n})$.
The static disorder is introduced by taking the site energies, 
$\epsilon_n$, to be independent,
identically distributed Gaussian random variables characterized by 
their variance $\sigma_n^2=\overline{\epsilon_n \epsilon_n}$. 
The overline is used throughout to denote the average over static disorder.
We assume that each site is independently coupled to its own phonon bath 
in the local basis. 
Thus $H_{\rm b} = \sum_{nk}\omega_{nk} b^{\dagger}_{nk}b_{nk}$ and 
$H_{\rm sb} = \sum_{nk}g_{nk}|n\rangle \langle n| (b^{\dagger}_{nk} + b_{nk}) $, 
where $\omega_{nk}$ and $b^{\dagger}_{nk}$($b_{nk}$) are the frequency 
and the creation (annihilation) operator of the $k$-th mode of the bath 
attached to site $n$ with coupling strength $g_{nk}$, respectively. 

Applying the polaron method to study the dynamics of open quantum systems 
was first proposed by Grover and Silbey~\cite{Grover2003}. This approach has gained a renewed attention 
due to the recent interest in light harvesting systems~\cite{McCutcheon2011a, Jang2008} and
has recently been extended to study non-equilibrium quantum transport~\cite{wang2014}. 
In this work, we will use a variant of the polaron based 
master equation which has the same structure as the popular Redfield equation. 

In the polaron technique, the unitary transformation operator, 
$\mbox{e}^{S}=\mbox{e}^{\sum_{nk} \frac{g_{nk}}{\omega_{nk}}|n\rangle \langle n|
(b^\dagger_{nk} - b_{nk}) }$, is applied to the total Hamiltonian 
\begin{equation} \po{H}_{\rm tot} = \mbox{e}^{S} H_{\rm tot} \mbox{e}^{-S} =
   \po{H}_{\rm s}+\tilde{H}_{\rm b}+\po{H}_{\rm sb},
\end{equation}
where $\po{H}_{\rm s} =\sum_n\epsilon_{n}|n\rangle \langle n| +   \sum_{m\neq n}\NormConst_{mn} J_{mn} |m\rangle \langle n|$, 
  $\po{H}_{\rm sb}  =  \sum_{n\neq m} J_{mn} |m\rangle \langle n| V_{mn}$ and 
  $ \po{H}_{\rm b} =  H_{\rm b} = \sum_{nk}\omega_{nk} b^{\dagger}_{nk}b_{nk}$.  
The electronic coupling is renormalized by a constant,  
$\NormConst_{mn}=
\mbox{e}^{-\frac{1}{2}\sum_{k}\left[\frac{g_{mk}^2}{\omega_{mk}^2}\coth(\beta
\omega_{mk}/2)+ \frac{g_{nk}^2}{\omega_{nk}^2}  \coth(\beta
\omega_{nk}/2)\right] }$, with the inverse thermal energy $\beta=1/k_B T$. 
The bath coupling operator now becomes $V_{mn} =
\mbox{e}^{\sum_{k}\frac{g_{mk}}{\omega_{mk}}(b^{\dagger}_{mk} - b_{mk} )}
\mbox{e}^{-\sum_{k}\frac{g_{nk}}{\omega_{nk}}(b^{\dagger}_{nk} - b_{nk} )}
-\NormConst_{mn}$, 
and is constructed such that its thermal average
is zero, i.e. $\mbox{tr}_{\rm b}[V_{mn} \mbox{e}^{-\beta H_{\rm b}}]=0$. 
Additionally, we assume the coupling
constants are identical across all sites $g_{nk}=g_k$ and a super-Ohmic spectral density, 
$J(\omega)=\pi\sum_k g_{k}^2\delta(\omega - \omega_k) = \gamma \omega^3 \mbox{e}^{-\omega /\omega_c}$ 
where $\gamma$ is the dissipation strength and $\omega_c$ is the cut-off frequency.

A detailed derivation of \PME\ is given in Appendix A, here we only summarize the main results. 
A perturbation approximation is applied 
in terms of the transformed system-bath coupling 
leading to a \PME\ for the transformed reduced density matrix, $\po{\rho}_s$:
\begin{eqnarray} 
   \frac{d\po{\rho}_{\nu \nu}(t)}{dt} &=&  \sum_{\nu'} R_{\nu \nu, \nu' \nu'} 
   \po{\rho}_{\nu' \nu'}(t); \\ 
   \frac{d \po{\rho}_{\mu \nu}(t)}{dt} &=&   (- \iu \,\omega_{\nu \mu} 
   + R_{\mu \nu, \mu\nu}) \po{\rho}_{\mu \nu}(t), \,\,\,\,\, 
   \mbox{$\nu \neq \mu$},  
\end{eqnarray} 
where the Markov and secular approximations have also been employed. 
The Greek indices denote the
eigenstates of the polaron transformed system Hamiltonian, 
i.e.    
$\po{H}_{\rm s} |\mu\rangle = \po{E}_\mu |\mu\rangle$ and 
$\omega_{\mu\nu} = \po{E}_\mu - \po{E}_\nu $.
The Redfield tensor,
$R_{\mu\nu,\mu'\nu'}$, describes the phonon-induced relaxation and can be
expressed as
\begin{align} 
   R_{\mu\nu,\mu'\nu'} &=  \Gamma_{\nu'\nu,\mu\mu'} +
   \Gamma_{\mu'\mu,\nu\nu'}^{*}
    -\delta_{\nu\nu'}\sum_{\kappa}
   \Gamma_{\mu\kappa,\kappa\mu'} -\delta_{\mu\mu'} \sum_{\kappa}
   \Gamma_{\nu\kappa,\kappa\nu'}^{*};\\
 \Gamma_{\mu\nu,\mu'\nu'}  &=  \sum_{mnm'n'}
   J_{mn}J_{m'n'}\langle \mu | m \rangle \langle n | \nu \rangle \langle \mu' |
   m' \rangle \langle n' | \nu' \rangle
   K_{mn,m'n'}(\omega_{\nu'\mu'})
   \;, \label{eq:ptre_rates}
\end{align}
where $K_{mn,m'n'}(\omega)$ is the half-Fourier transform of the bath
correlation function
\begin{equation} K_{mn,m'n'}(\omega) = \int^{\infty}_{0} \mbox{e}^{\iu \omega
   t} \langle V_{mn}(t)V_{m'n'}(0)\rangle_{H_{\rm b}} dt \;,  
\end{equation}
and $\langle V_{mn}(t)V_{m'n'}(0)\rangle_{H_{\rm b}} = \mbox{tr}_{\rm b}[\mbox{e}^{-\beta H_{\rm b}} V_{mn}(t)V_{m'n'}(0) ]/  \mbox{tr}_{\rm b}[\mbox{e}^{-\beta H_{\rm b}}]$.
Since the system is disordered and we are mainly interested in the long time
dynamics, the Markov and secular approximations do not incur a significant
loss of accuracy. 
Comparison of the dynamics computed with and without these approximations 
shows little discrepancy (see Appendix A).
It should be noted that the transformed reduced density matrix, $\po{\rho}_s$,  
is different from the reduced density matrix in the original frame, $\rho_s$. 
However, for the transport properties studied here, only the population 
dynamics is needed which is invariant under the polaron transformation since 
$\po{\rho}_{nn}(t) = \rho_{nn}(t)$. 

\textit{Diffusion} - In the presence of both disorder and dissipation, we find
empirically that after an initial transient time that is approximately 
proportional to $J^3 \beta/\gamma$, the mean square displacement,
$\overline{\langle R^2(t) \rangle}= \overline{\sum_{n} n^2 \rho_{nn}(t)}$, 
grows linearly with time, where the origin is defined such that $\langle R^2(0) \rangle=0$. 
The diffusion constant, $D$, can then be defined as 
$\lim_{t\rightarrow \infty} \overline{ \langle R^2(t) \rangle} = 2D t$. 
The electronic coupling $J=1$ sets the energy scale 
and quantities throughout are implicitly stated in units of $J$. 
In the numerical simulations, we use a one-dimensional chain of 250-300
sites and average over 100-500 realizations of static disorder
sampled from a Gaussian distribution with variance $\sigma=1$.
The number of realizations needed for convergence  is highly dependent 
on the temperature; more samples are needed in the low temperature regime.

\section{Results} 
\label{section:results}
\begin{figure}  
   \includegraphics*[width=6in]{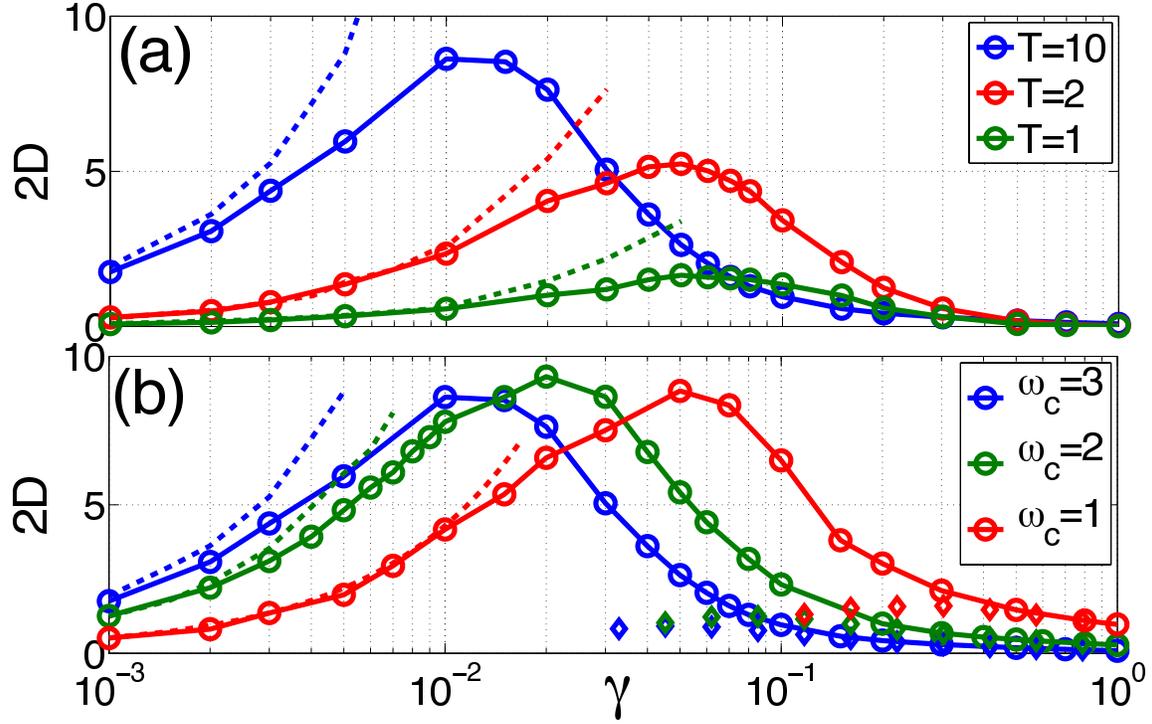}
   \caption{
      The diffusion constant as a function of the dissipation 
      strength, $\gamma$. 
      The dashed lines display the corresponding results from the SRE, 
      while diamond symbols depict the results of
      the FGR rates given in \autoref{eqn:fermi}. 
      (a) Results for different temperatures and a fast bath $\omega_c=3$. 
      (b) Results for different cut-off frequencies and $T=10$. 
   } \label{FIG:gamma} 
\end{figure}

We first investigate the effect of the dissipation strength,
$\gamma$, on the diffusion constant in \autoref{FIG:gamma}. 
A non-monotonic dependence of $D$ as a function of $\gamma$ is observed,
consistent with the previous studies using the Haken-Strobl 
model~\cite{Moix2013a}.
Without coupling to the bath, there is no macroscopic transport since 
all the wavefunctions in the one-dimensional disordered system are Anderson
localized. 
Introducing dissipation destroys the phase coherence that gives rise to 
localization, allowing for transport to occur. 
Therefore in the weak coupling regime, increasing $\gamma$
leads to faster transport 
as is readily apparent from the SRE rates in \autoref{eq:ptre_rates}, 
and thus $D$ increases linearly with $\gamma$.
In the opposite regime of strong coupling, the coherence generated between 
sites is quickly destroyed and the quantum transport reduces to
a classical hopping dynamics between neighboring sites. 
In this regime, the dissipation strength effectively acts as classical friction 
that impedes the transport~\cite{Cao2009} leading $D$ to behave as a 
decreasing function of $\gamma$ as is observed in \autoref{FIG:gamma}. 
The interplay between static disorder and dissipation thus
gives rise to an optimal dissipation strength for transport. 
In \autoref{FIG:gamma} (a), it is seen that the maximal diffusive rate 
both increases and shifts to smaller coupling strengths as the temperature 
increases since thermal fluctuations also assist the quantum system to
overcome the localization barriers in the weak coupling regime. 
For comparison, we also include the results from the SRE 
in the weak coupling regime. 
For small $\gamma$ and $T$, the SRE provides a
reliable description of the transport properties, but starts to breakdown as 
$\gamma$ (or $T$) increases leading to an unphysical 
$D \propto \gamma$ dependence. 
The breakdown of the SRE has been discussed by 
Ishizaki and Fleming for a two-site model~\cite{Ishizaki2009} and by 
Wu \textit{et al.} for FMO~\cite{Wu2010}.

\autoref{FIG:gamma} (b) depicts $D$ as a function of
$\gamma$ for different bath cut-off frequencies. 
It is found that the large $\gamma$ scaling of $D$ is highly dependent 
on the relaxation time of the bath. 
For a fast bath, the rates decrease approximately as $1/\gamma$ 
consistent with our previous analysis of the Haken-Strobl 
model~\cite{Moix2013a}.
However, as the bath frequency decreases,
a transition from the $1/\gamma$ dependence to
$1/\sqrt{\gamma}$ dependence is observed. 
This can be rationalized by noting that 
in the high temperature and strong damping regime, 
the dynamics are incoherent and can be described by classical hopping
between nearest neighbors.
Then the hopping rate between sites $m$ and $n$ is accurately 
determined from FGR:
\begin{equation}
   k_F (\Delta_{mn}) = J_{mn}^2\kappa_{mn}^2 \,\mbox{Re} \int^\infty_{-\infty} dt \,\,
   \mbox{e}^{\iu \Delta_{mn} t} [\mbox{e}^{  g(t)}  -1], 
   \label{eqn:fermi} 
\end{equation} 
and
\begin{equation} g(t) =  2\int^\infty_0
  \frac{d\omega}{\pi}\frac{J(\omega)}{\omega^2}\Big[ \coth\frac{\beta
     \omega }{2}  \cos \omega t - \iu \sin \omega t\Big],
\end{equation}
where $\Delta_{mn} = \epsilon_m - \epsilon_n$ is the activation barrier, and 
$J_{mn}$ is the electronic coupling.  
In the slow bath limit 
the above expression reduces to the Marcus rate $k_M(\Delta) \approx \frac{\pi}{2} J^2 \sqrt{\frac{\beta}{\gamma \omega_c^3}} \mbox{e}^{-\frac{ \beta( \pi \Delta - 4\gamma \omega_c^3)^2}{ 16 \pi \gamma \omega_c^3}}$ 
which captures the correct $1/\sqrt{\gamma T}$ dependence of the rate.
Defining the energy transfer time as the inverse of the rate, 
$\tau_F(\Delta) = 1/k_F(\Delta)$, static disorder can be introduced by
averaging $\tau_F(\Delta)$ over the Gaussian distribution of static 
disorder: $\overline{ \tau_F} = \int P(\Delta) \tau_F(\Delta) d\Delta$ 
where $P(\Delta) = \frac{1}{\sigma' \sqrt{2\pi}}\mbox{e}^{-\Delta^2/2\sigma'^2}$
and $\sigma'^2 = \overline{\Delta_{mn}^2}=2\sigma^2$.
The disorder-averaged golden rule rate can then be obtained using 
$\overline{ k_F} =1/\overline{\tau_F}$ and is plotted in \autoref{FIG:gamma} (b). 
While it is seen to capture the correct scaling of $D$ in the overdamped 
regime, it significantly underestimates the transport in 
the small and intermediate damping regimes. 
As the dynamics becomes more coherent, the classical
hopping rate between sites provides a qualitatively incorrect 
description of the transport.  
In the sufficiently weak dephasing regime, 
the dynamics from the \PME\ reduce to those 
of the standard SRE  (dashed lines in \autoref{FIG:gamma} (b)).

\begin{figure}  
   \includegraphics*[width=6in]{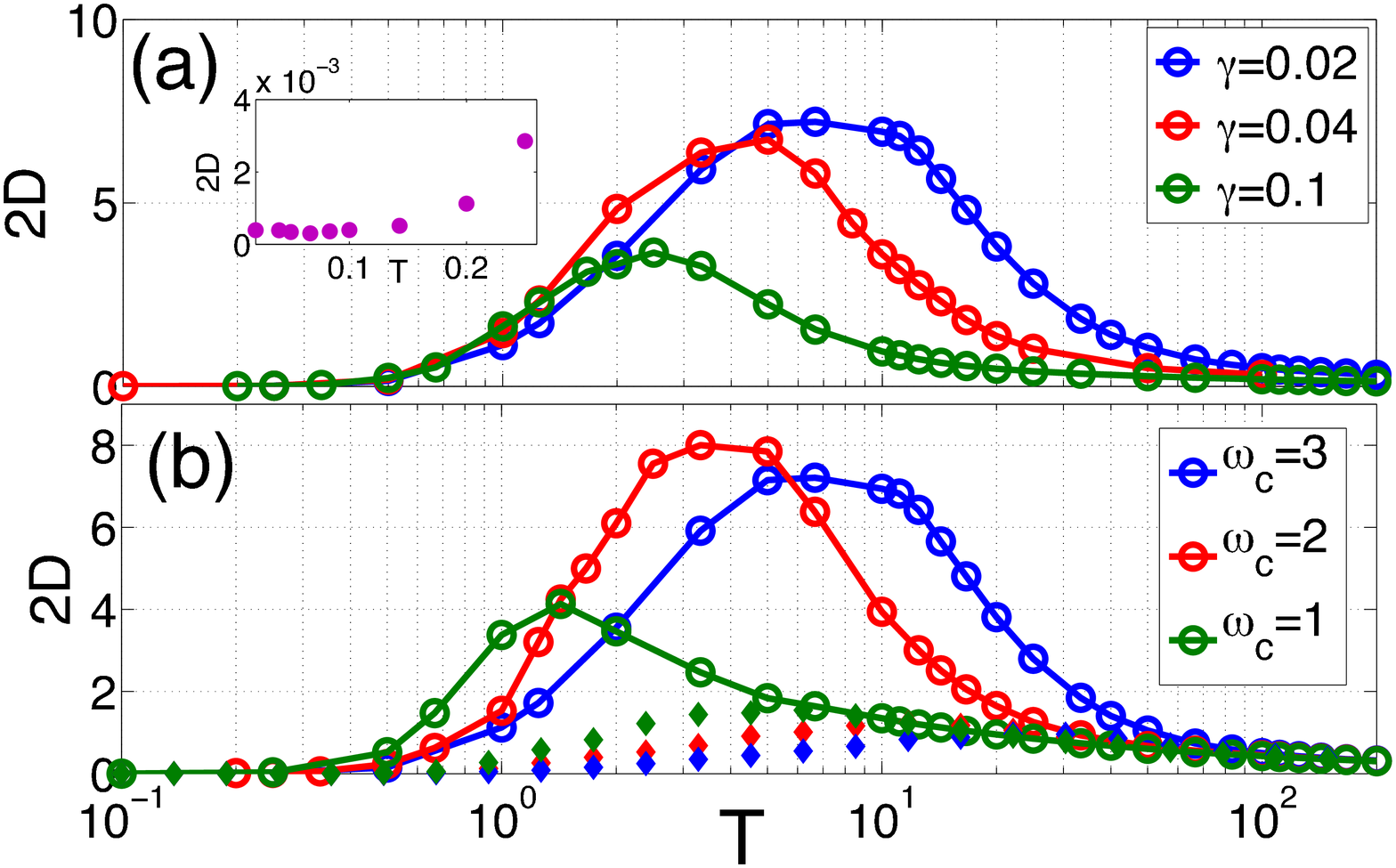}
   \caption{The diffusion constant as a function of temperature. 
      (a) Results for different values of dissipation strength and a fast bath
      $\omega_c=3$.  
      The inset shows the diffusion constant calculated with the SRE near zero 
      temperature and $\gamma=0.01$.
      (b) Results for different cut-off frequencies and a constant
      reorganization energy of $\int^\infty_0 \frac{J(\omega)}{\omega} = 1.08$.
      The diamond symbols depict the results of the FGR rates 
      as given by \autoref{eqn:fermi}. 
   } \label{FIG:temperature} 
\end{figure}

While the $\gamma$ dependence of $D$ provides many physical insights, 
the temperature dependence is more experimentally accessible,
and is presented in \autoref{FIG:temperature}. 
Similar to the $\gamma$ dependence, $D$ exhibits a
non-monotonic dependence of $T$ and the high $T$ scaling is 
sensitive to the cut-off frequency of the bath, as shown in 
\autoref{FIG:temperature} (b). 
At high temperature, we observe $D$ decreases as $1/\sqrt{T}$ 
for a slow bath as predicted by the Marcus theory,
while for a fast bath, we recover the Haken-Strobl scaling of $1/T$.
The system-bath coupling strengths shown in \autoref{FIG:temperature} (a)
lie to the right of the maxima in \autoref{FIG:gamma}.
Hence $D$ decreases as $\gamma$ increases in the high
temperature regime.
In the opposite low temperature regime, the intermediate coupling results shown
in \autoref{FIG:temperature} (a) are beyond the reach of the SRE. 
Thus $D$ does not increase at a rate proportional to $\gamma$ as might 
be expected. 
However, the results here also do not agree with the Marcus formula
where one would expect the transport rate to decrease as $1/\sqrt{\gamma T}$,
but instead the diffusion constants are nearly independent of $\gamma$.
This low temperature, intermediate coupling regime is not adequately 
described by either of the perturbative methods.  
At zero temperature, quantum fluctuations from the thermal environment 
are still present to destroy the Anderson localization 
and allow for transport to occur, albeit at a very slow rate. 
This leads to a small but finite value of $D$ as seen
in the inset of \autoref{FIG:temperature} (a).

\begin{figure}  
   \includegraphics*[width=6in]{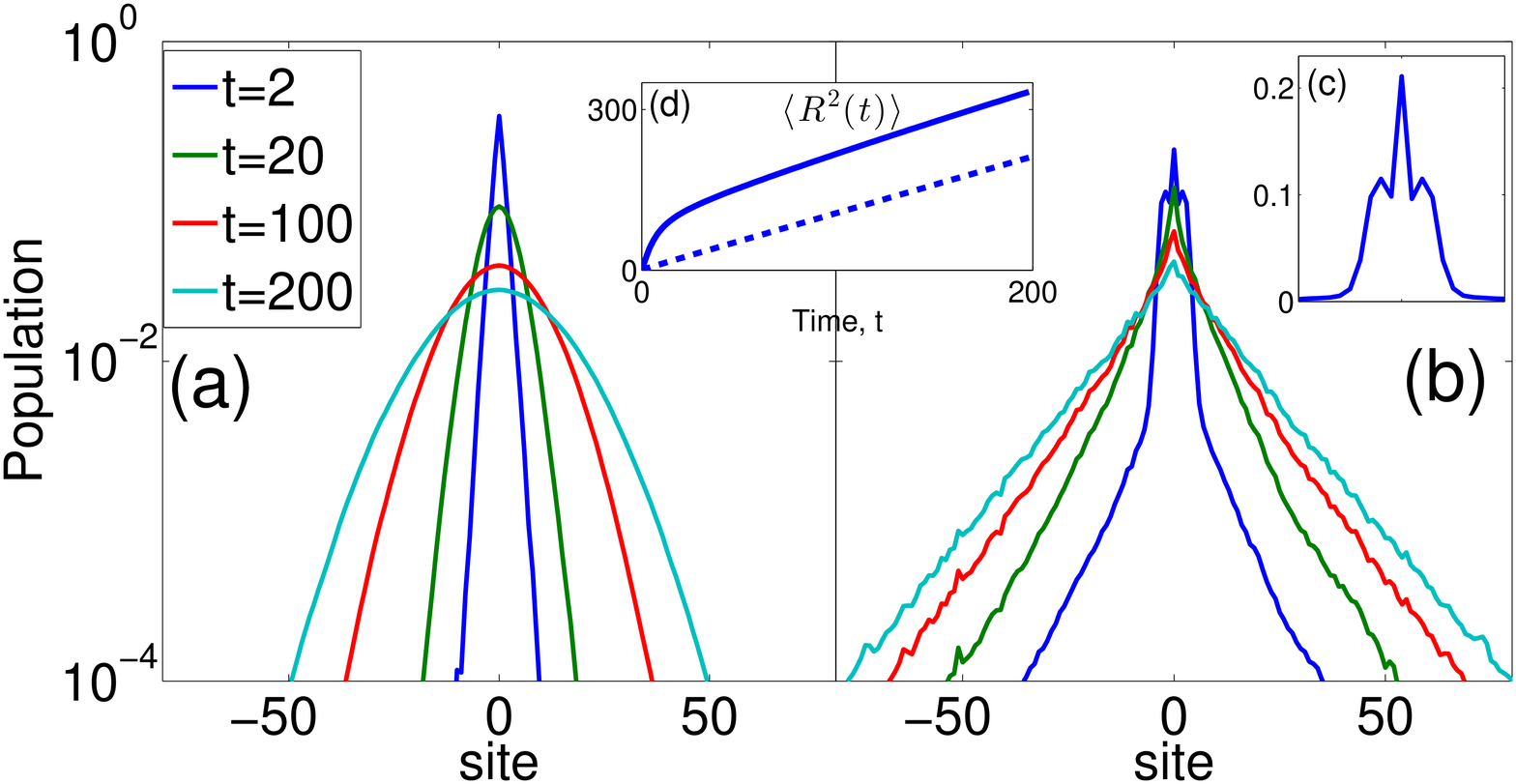}
   \caption{Time evolution of the probability distribution profiles 
      for $\gamma=0.02$, $\omega_c=3$ and (a) high temperature, $T=50$ 
      or (b) low temperature $T=1$. 
      The diffusion constant is $2D\approx 1.1$ in both cases.
      The inset (c) shows the low temperature wavelike population
      profile at $t=2$ in linear scale.
      The inset (d) shows the respective mean-squared displacements 
      $\langle R^2(t) \rangle$ for $T=1$ (solid) and $T=50$ (dashed). 
   }
   \label{FIG:dynamics} 
\end{figure}

In addition to the steady state transport properties, it is also useful 
to explore the dynamics of noisy, disordered systems.  
\autoref{FIG:dynamics} displays the average population probability 
distribution at high and low temperatures for an initial excitation 
located at the center of the disordered chain. 
The temperatures are selected such that the diffusion
constants in \autoref{FIG:dynamics} (a) and (b) are approximately the same, $2D \approx 1.1$. 
In the high temperature case, 
the coherence is quickly destroyed by dissipation, no wavelike
motion is observed in the time scale plotted. 
While the population distribution appears exponential at short times
--which is a signature of Anderson localization--
the exponential behavior quickly transitions to a Gaussian profile  
indicating the onset of the diffusive regime. 
The population dynamics at low temperature in \autoref{FIG:dynamics} (b)
is qualitatively different.
At short times, the distribution near the center of the chain 
(\autoref{FIG:dynamics} (c)) displays 
wavelike motion characteristic of free-particle dynamics
while the tails decay exponentially. 
The wavelike motion disappears at intermediate times but the localization peak
near the center persists. 
Although the exponential tail eventually disappears,
the transition of the population distribution to a Gaussian form
is slow and takes place long after a reliable estimate of $D$ 
can been obtained
$\left(\overline{\langle R^2 \rangle} \propto t\right)$, 
as shown in inset of the \autoref{FIG:dynamics} (d).

\section{Applications}
\label{section:applications}
 It is interesting to compare 
estimates of the transport properties in real material systems
from the \PME\ with the approximate FGR and SRE rates that 
are often assumed to hold. 
For example, predictions of the charge mobility,
$\mu =\frac{e D}{k_{\rm B} T}$, of several commonly used organic
semiconductors are presented in \autoref{tab:hresult}.
The parameters are taken from Ref.~\cite{Stehr2011} and 
we use the directions with the largest electronic couplings. 
Despite using a vastly simplified model, the mobility calculated with 
the \PME\ is in reasonable agreement with the available experimental values,
while the SRE usually leads to an substantial overestimation
of the mobility and the commonly used FGR generally leads to a 
significant underestimation because of the neglect of quantum coherence.

\section{Conclusion} 
\label{section:conclusion}
We have developed a polaron-transformed 
Redfield equation to systematically 
study the transport properties of disordered systems
in the presence of quantum phonon modes, and established scaling relations for 
the diffusion coefficients at both limits of the temperature and 
system-bath coupling strength.
The results presented here constitute one of the first 
studies of quantum transport in extended disordered systems over 
the complete range of phonon bath parameters. 
The \PME\ provides a general framework to establish a unified understanding 
of the transport properties of a wide variety of systems including 
light-harvesting complexes, organic photovoltaics, conducting polymers 
and J-aggregate thin films. 

\section{acknowledgement}
This work was supported by the NSF (Grant No.~CHE-1112825) and DARPA 
(Grant No.~N99001-10-1-4063). 
C.~K.~Lee acknowledges support by the Ministry of Education
(MOE) and National Research Foundation of Singapore.  J.~Moix has been
supported by the Center for Excitonics, an Energy Frontier Research Center
funded by the US Department of Energy, Office of Science, Office of Basic
Energy Sciences under Award No.~DE-SC0001088. 

\appendix

\section{Derivation of the Polaron-Transformed Master Equation}
The total Hamiltonian of the system and the bath is 
\begin{eqnarray}
		H_{tot} &=& H_s + H_b + H_{sb}, \\ \nonumber
		         &=& \sum_{n} \epsilon_{n}|n\rangle \langle n| + \sum_{m \neq n} J_{mn} |m\rangle \langle n| + \sum_{n} \sum_{k} \omega_{nk} b^\dagger_{nk} b_{nk} + \sum_{n} \sum_{k} g_{nk} |n\rangle \langle n| (b^\dagger_{nk} + b_{nk}),
\end{eqnarray}
where $|n \rangle$ denotes the site basis, $\epsilon_n$ is the site energy, and $J_{nm}$ is the electronic coupling between site $n$ and site $m$. 
Each site is independently coupled to its own phonon bath in the local basis. The variable $\omega_{nk}$ and operator $b^{\dagger}_{nk}$($b_{nk}$) are the frequency 
and the creation (annihilation) operator of the $k$-th mode of the bath 
attached to site $n$ with coupling strength $g_{nk}$, respectively. 

Applying the polaron transformation, $\mbox{e}^S=\mbox{e}^{\sum_{nk} \frac{g_{nk}}{\omega_{nk}} |n\rangle \langle n|(b^\dagger_{nk} - b_{nk}) }$, to the total Hamiltonian, we obtain
\begin{eqnarray}
		\po{H}_{tot} &=& \mbox{e}^S H_{tot} \mbox{e}^{-S} = \po{H}_{s} + \po{H}_{b} + \po{H}_{sb}; \\
		\po{H}_{s}   &=& \sum_{n} \epsilon_n |n\rangle\langle n| + \sum_{m \neq n} \kappa_{mn}J_{mn} |m\rangle \langle n|;\\
		\po{H}_{b}   &=& H_b = \sum_{n} \sum_{k} \omega_{nk} b^\dagger_{nk} b_{nk};\\
		\po{H}_{sb}  &=& \sum_{m \neq n}J_{mn} |m\rangle \langle n|V_{mn},
\end{eqnarray}
where the electronic coupling is renormalized by the constant,  
$\kappa_{mn}=
\mbox{e}^{-\frac{1}{2}\sum_{k}\frac{g_{mk}^2}{\omega_{mk}^2}\coth(\beta
\omega_{mk}/2)  -\frac{1}{2}\sum_{k}\frac{g_{nk}^2}{\omega_{nk}^2}  \coth(\beta
\omega_{nk}/2)}$ and $\beta = 1/k_B T$. Tildes are used to denote operators
in the polaron frame.
The bath coupling operator now becomes $V_{mn} =
\mbox{e}^{\sum_{k}\frac{g_{mk}}{\omega_{mk}}(b^{\dagger}_{mk} - b_{mk} )}
\mbox{e}^{-\sum_{k}\frac{g_{nk}}{\omega_{nk}}(b^{\dagger}_{nk} - b_{nk} )}
-\kappa_{mn}$. The bath coupling term is constructed such that its thermal average
is zero, $\mbox{tr}_{b}[V_{mn} \mbox{e}^{-\beta H_b}]=0$. 
Assuming the coupling constants are identical across all sites $g_{nk}=g_k$ and
introducing the spectral density $J(\omega)=\pi\sum_k g_{k}^2 \delta(\omega -
\omega_k)$, the renormalization constant can then be written as $\kappa=\kappa_{mn}=
\mbox{e}^{-\int^{\infty}_{0} \frac{d \omega}{\pi}
\frac{J(\omega)}{\omega^2}\coth(\beta \omega/2) }$. Here we use a
super Ohmic spectral density $J(\omega) = \gamma \omega^3 \mbox{e}^{-\omega
/\omega_c}$ where $\gamma$ is the dissipation strength and $\omega_c$ is the
cut-off frequency.

To derive the master equation, let us first introduce the Hubbard operator $X_{\nu\mu}= |\nu\rangle \langle \mu|$, where $|\nu \rangle$ is the eigenstate of the transformed system Hamiltonian: $\po{H_s}|\nu \rangle= E_{\nu}|\nu \rangle $. The system reduced density matrix element can be conveniently obtained using the relation $\rho_{\mu\nu}(t)=\mbox{tr}_{s+b}[\rho_{s+b}(0) X_{\nu\mu}(t)]$, as will be done later. The Heisenberg equation of the Hubbard operator is given by

\begin{eqnarray}
		\frac{d X_{\nu\mu}(t)}{dt} &=& \iu[ \po{H}_{0}(t), X_{\nu\mu}(t)] + \iu [\po{H}_{sb}(t), X_{\nu\mu}(t)], \label{eqn:heisenberg}
\end{eqnarray}
where $\po{H}_0=\po{H}_s+\po{H}_b$ is the free Hamiltonian. We can write the second term as 
\begin{eqnarray}
   [\po{H}_{sb}(t), X_{\nu\mu}(t)] = U^\dagger(t) \,\,[\po{H}_{sb}, X_{\nu\mu}]\,\, U(t), \label{eqn:heisenberg2}
\end{eqnarray}
where the evolution operator is $U(t)= \mbox{e}^{- \iu \po{H}_{tot} t}$. 
We then use Kubo's identity~\cite{kubo} to expand $U(t)$ perturbatively 
in terms of $\po{H}_{sb}$:
\begin{eqnarray}
		U(t) \approx \mbox{e}^{-\iu \po{H}_0 t} \left[1 - \iu \int^t_0 ds\: \intpic{\po{H}}_{sb}(s)\right],
\end{eqnarray}
where hats over the operators are used to denote operators in the interaction
picture, $\intpic{O}(t) = \mbox{e}^{\iu {\po H }_0 t} O \mbox{e}^{-\iu \po{H}_0
t}$. Inserting the expansion into \autoref{eqn:heisenberg2} and keeping terms
up to second order in $\po{H}_{sb}$, the Heisenberg equation, \autoref{eqn:heisenberg}, becomes
\begin{eqnarray}
		\frac{dX_{\nu\mu}(t)}{dt} = \iu [\po{H}_0(t), X_{\nu\mu}(t)] + \iu [\widehat{\po{H}}_{sb}(t), \intpic{X}_{\nu\mu}(t)] -\int^t_0 ds \,[ \intpic{\po{H}}_{sb}(s),\,[\intpic{\po{H}}_{sb}(t), \intpic{X}_{\nu\mu}(t)] ]\label{eqn:heisenberg3}
\end{eqnarray}

We multiply the initial condition, $\po{\rho}_{s+b}(0)$, to the RHS of
\autoref{eqn:heisenberg3} and perform a total trace of both the system and bath, obtain an equation governing the dynamics of the system reduced density
matrix elements 
\begin{eqnarray}
    \frac{d\po \rho_{\mu\nu}(t)}{dt} 
       &=& - \iu\,\omega_{\mu\nu}\po\rho_{\mu\nu}(t)  
       -\int^t_0 ds \,\, \mbox{tr}_{s+b}\Big(
                   [\intpic{\po{H}}_{SB}(s),\,[\intpic{\po{H}}_{sb}(t),
                      \po{X}_{\nu\mu}(t)] \,] \po{\rho}_{s+b}(0) \Big),
                      \label{eqn:master}
\end{eqnarray}
where $\omega_{\mu\nu} = E_\mu- E_\nu$. 
Assuming factorized initial conditions,
$\po{\rho}_{s+b}(0) = \po{\rho}_s(0) \otimes \frac{\mbox{e}^{-\beta
H_b}}{\mbox{tr}[\mbox{e}^{-\beta H_b}]}$, and substituting the expression of
$\po{H}_{sb}$ into \autoref{eqn:master}, we finally have the master equation
after some
 manipulations
\begin{eqnarray}
	\frac{d\tilde{\rho}_{\mu\nu}(t)}{dt} = - \iu \omega_{\mu\nu} \tilde{\rho}_{\mu\nu}(t) + \sum_{\mu'\nu'}R_{\mu\nu,\mu'\nu'}(t) \tilde{\rho}_{\mu'\nu'}(t),
\end{eqnarray}
where the Redfield tensor, $R_{\mu\nu,\mu'\nu'}(t)$, describes the
phonon-induced relaxation. 
It can be expressed as
\begin{eqnarray}
	R_{\mu\nu,\mu'\nu'}(t) =  \Gamma_{\nu'\nu,\mu\mu'}(t) + \Gamma_{\mu'\mu,\nu\nu'}^{*}(t) -\delta_{\nu\nu'}\sum_{\kappa} \Gamma_{\mu\kappa,\kappa\mu'}(t) -\delta_{\mu\mu'} \sum_{\kappa} \Gamma_{\nu\kappa,\kappa\nu'}^{*}(t). \label{eqn:redfield}
\end{eqnarray}
The damping rates have the form
\begin{eqnarray}
	\Gamma_{\mu\nu,\mu'\nu'}(t)  =  \sum_{mnm'n'} J_{mn}J_{m'n'}\langle \mu | m \rangle \langle n | \nu \rangle \langle \mu' | m' \rangle \langle n' | \nu'\rangle K_{mn,m'n'}(\omega_{\nu'\mu'}, t),
\end{eqnarray}
where $K_{mn,m'n'}(\omega, t)$ is the integrated bath correlation function
\begin{eqnarray}
	K_{mn,m'n'}(\omega, t) = \int^{t}_{0} \mbox{e}^{\iu \omega t} \langle \intpic{V}_{mn}(t) \intpic{V}_{m'n'}(0)\rangle_{H_b} d\omega.
\end{eqnarray}
$ \langle \bullet \rangle_{H}$ denotes the average over the thermal state $\mbox{e}^{-\beta H}/  \mbox{tr}[\mbox{e}^{-\beta H}]$. Assuming a short bath correlation time, we can make the Markov approximation by taking the upper integration limit to infinity, making the damping rate a half-Fourier transform of the bath correlation function. The bath correlation function is given by \cite{Jang2011}
\begin{eqnarray}
	 \langle V_{mn}(t)V_{m'n'}(0)\rangle_{H_b}  = \kappa^2(\mbox{e}^{-\lambda_{mn,m'n'} \phi(t)}-1),
\end{eqnarray}
where $\lambda_{mn,m'n'}= \delta_{mm'} -\delta_{mn'} +\delta_{nn'}- \delta_{nm'}$ and
\begin{eqnarray}
		\phi(t)=\int^\infty_0 \frac{d\omega}{\pi}\frac{J(\omega)}{\omega^2}\Big[\cos(\omega t) \coth(\beta \omega/2) \,-\, \iu \sin(\omega t) \Big].
\end{eqnarray}

A few remarks are in place. 
First, the decoupled initial condition assumption
is generally not true in the polaron frame since $\po{\rho}_{s+b}(0)$ is
usually a complicated system-bath entangled state generated by the 
polaron transformation. 
This occurs even if the initial state in the original frame does factorize as
$\rho_{s+b}=\rho_s(0)\otimes \rho_b(0)$. 
Regardless, in most cases, the decoupled initial condition is only 
an approximation. 
However, Nazir \textit{et al.}~\cite{Kolli2011} has shown that the error 
incurred due to the initial condition is only significant at short times. 
In this work, we are mainly interested in the long time dynamics of the 
system. 
Therefore,  the accuracy of our results is not  considerably 
affected by the decoupled initial condition approximation.
Second, \autoref{eqn:redfield} has the same structure as the Redfield
equation commonly used in the magnetic resonance literature. The only
difference is that the damping tensor $\Gamma_{\mu\nu,\mu'\nu'}$ contains four
summations as opposed to two summations in the standard Redfield equation. 
In fact, the Redfield equation can be recovered by
following the same prescription as described above except without applying
the polaron transformation. 

Within the secular approximation, the evolution of the diagonal and off-diagonal density matrix elements are decoupled:
\begin{eqnarray}
		\frac{d\tilde{\rho}_{\nu \nu}(t)}{dt} &=&  \sum_{\nu'} R_{\nu \nu, \nu' \nu'} \rho_{\nu' \nu'}(t);  \label{eqn:secular_pop} \\
                \frac{d \tilde{\rho}_{\mu \nu}(t)}{dt} &=&   (- \iu
                \,\omega_{\nu \mu} + R_{\mu \nu, \mu\nu}) \tilde \rho_{\mu \nu}(t),
                \,\,\,\,\, \text{$\nu \neq \mu$}. \label{eqn:secular_coh}
\end{eqnarray}

\begin{figure}  
 \begin{center}
 \includegraphics[width=3.3in]{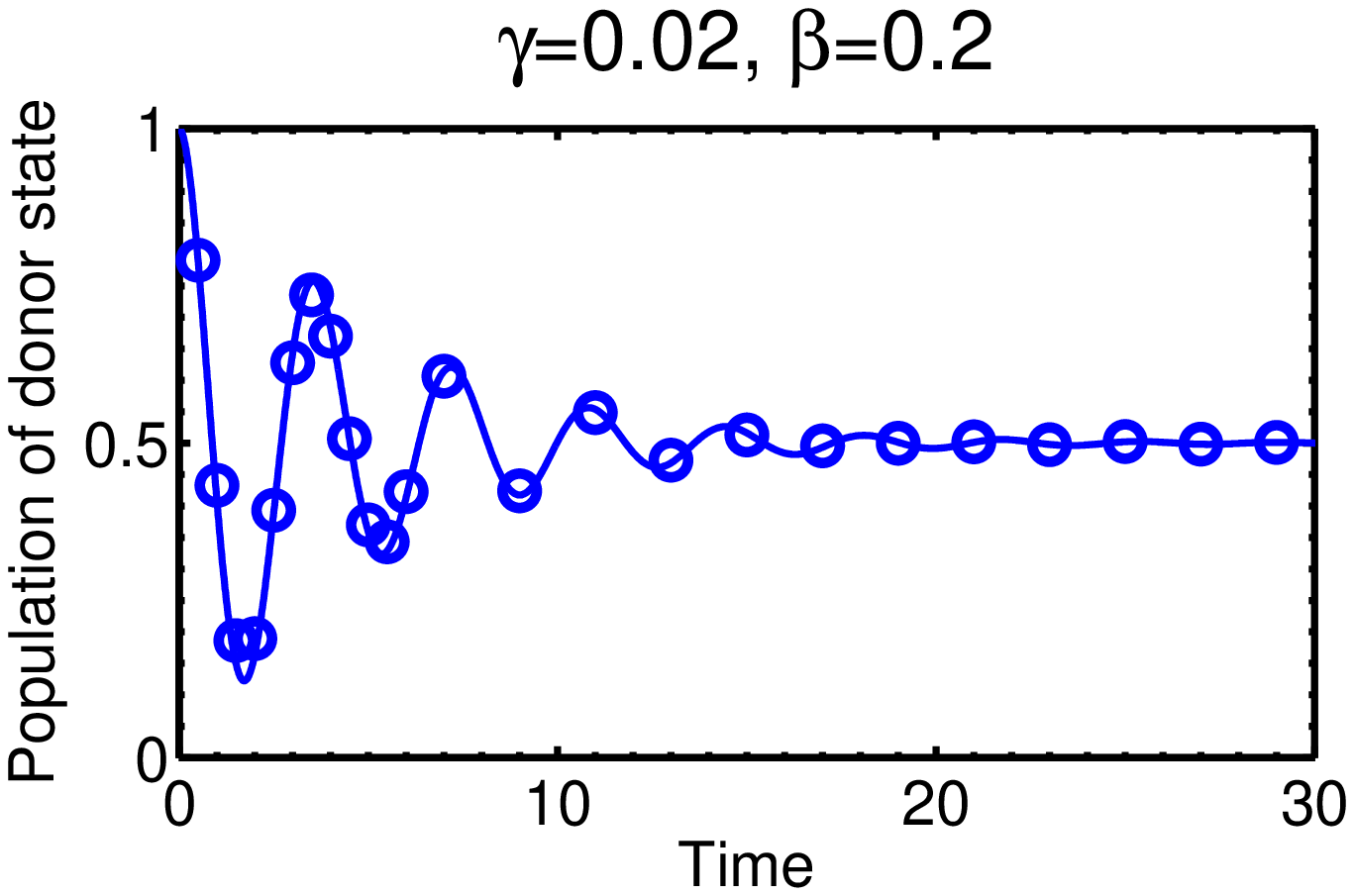} 
 \includegraphics[width=3.3in]{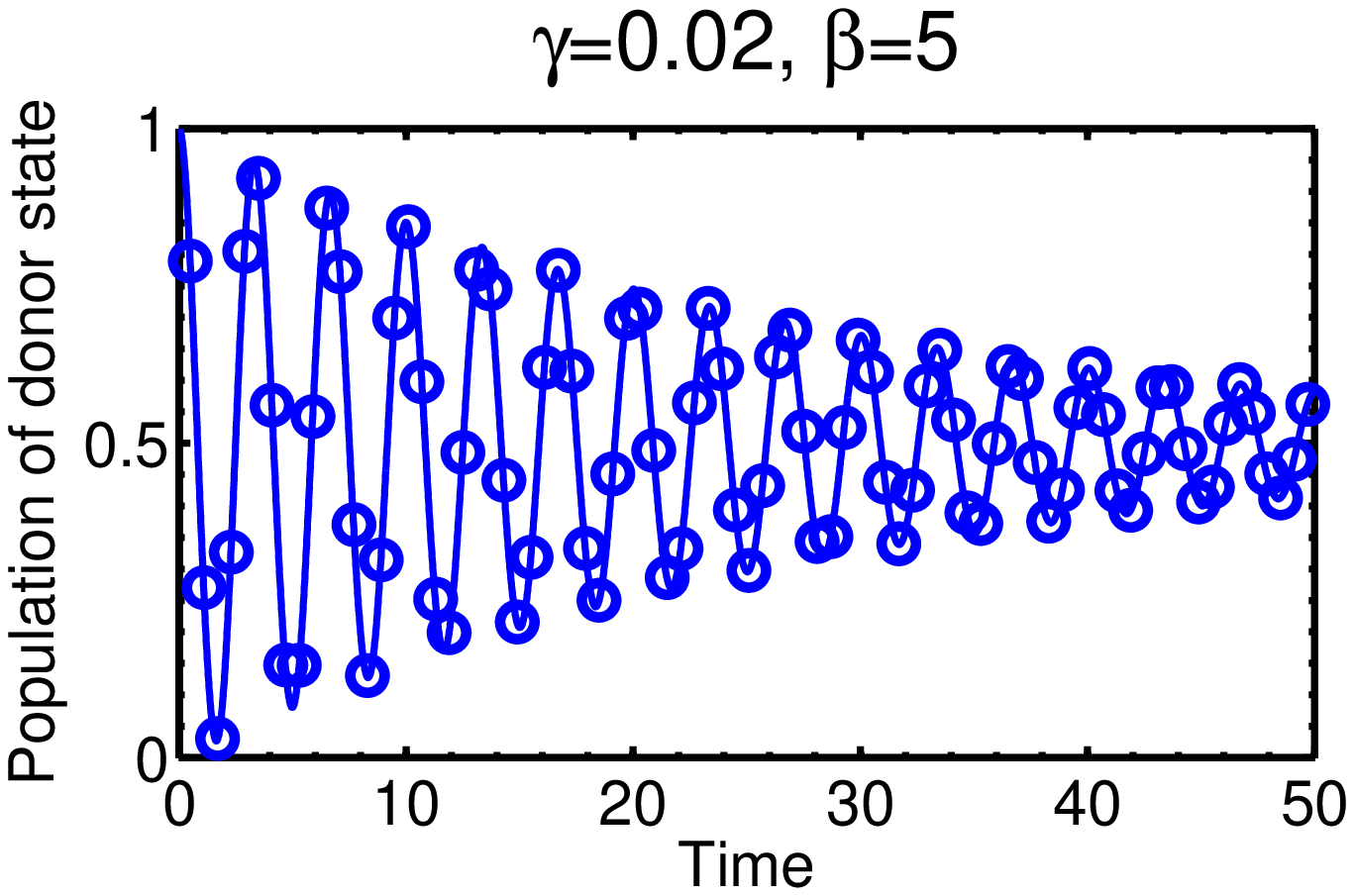}
 \includegraphics[width=3.3in]{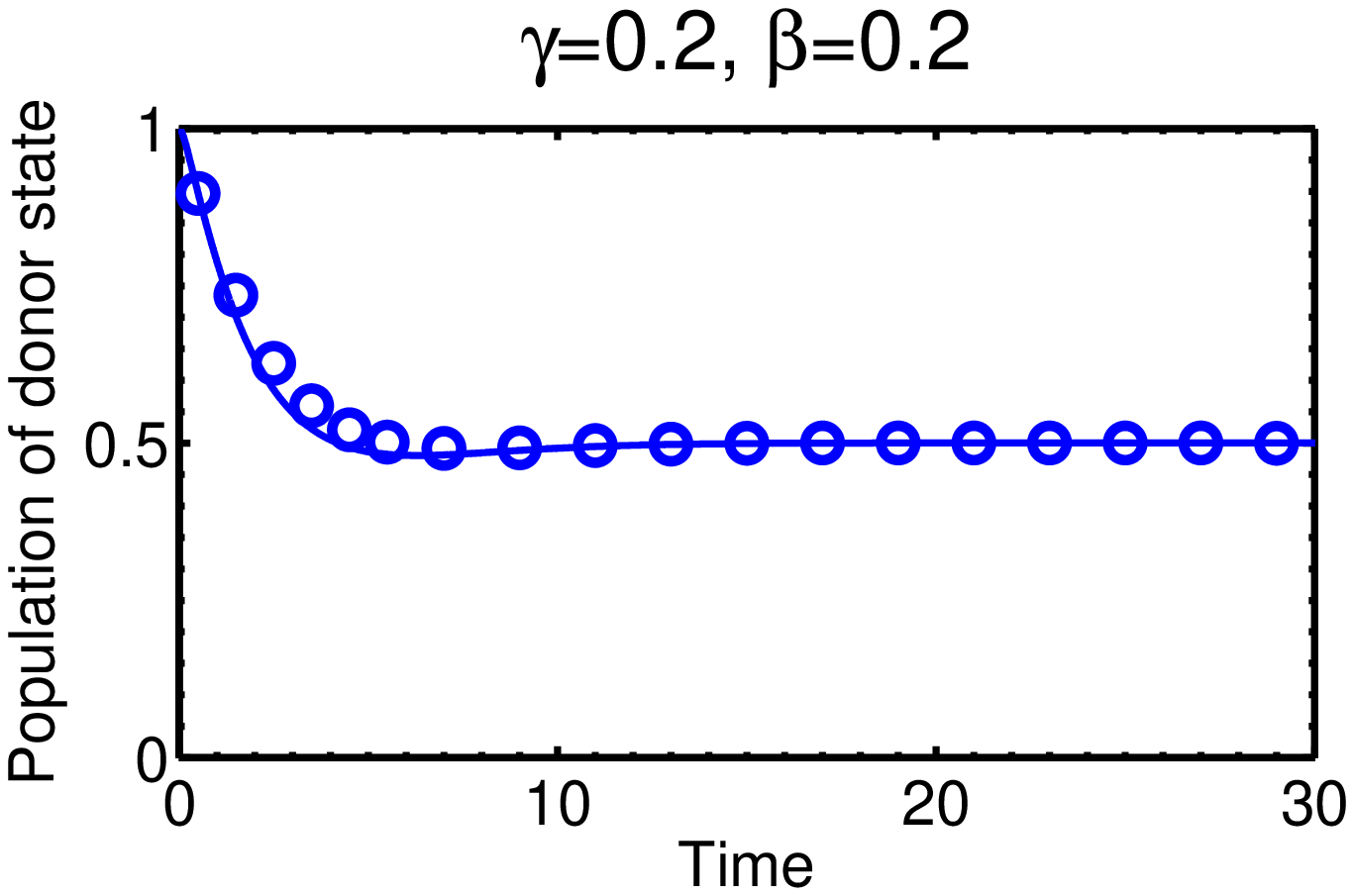} 
 \includegraphics[width=3.3in]{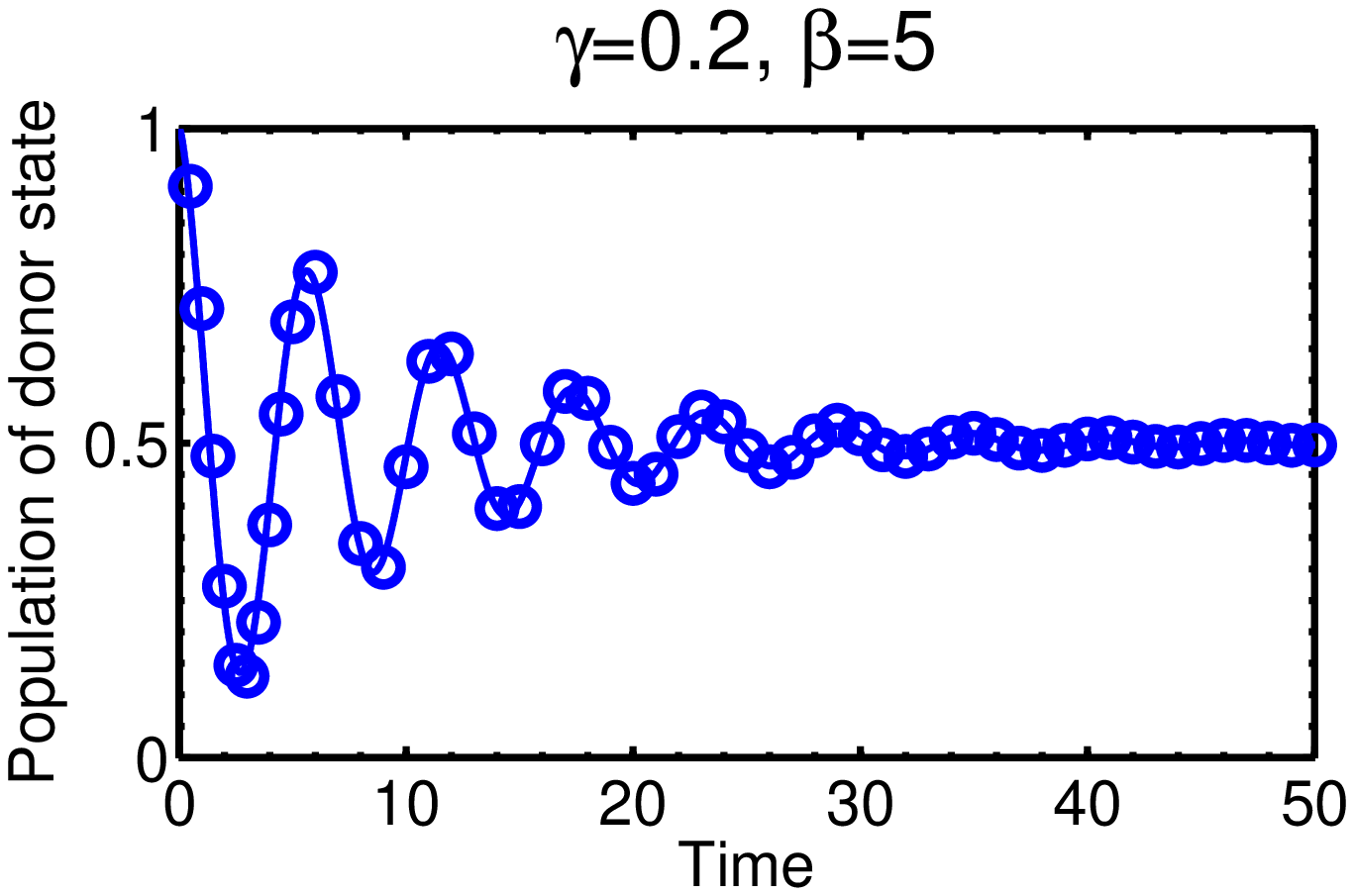} 
 \end{center}
   \caption{
      Time evolution of the population dynamics of the donor (site 1)
      calculated using the secular polaron master equation as given in
      \autoref{eqn:secular_pop} and \autoref{eqn:secular_coh} (symbols) and the time-convolutionless
      second-order polaron master equation used in Ref.~\cite{Zimanyi2012}
      (solid lines). The parameters used are $\epsilon_1=\epsilon_2=0$,
      $J_{12}=1$ and $\omega_c=3$.
   }
   \label{figure:comparison}
\end{figure}

We compare the results from the above secular polaron master equation 
with that of the more accurate time-convolutionless second-order polaron master
equation~\cite{Zimanyi2012} without the secular and Markov approximations for an 
unbiased two-site system. The results are plotted in
\autoref{figure:comparison}. It can the be seen that the results agree
remarkably well for different values of temperature and 
coupling strength.
This demonstrates that the secular and Markov approximations made here do not
incur a significant loss of accuracy in our results.
%

\section{Strong Damping Limit}
Here we explore the strong system-bath coupling limit of the \PME. 
In the strong coupling limit, the coherence is quickly destroyed by dissipation, 
thus we only need to consider the equations of motion of the population, 
i.e. \autoref{eqn:secular_pop}.
Additionally, $\kappa_{mn} \rightarrow 0$ for large $\gamma$, i.e.  
the eigenbasis of $\po{H}_s$ is also the site basis, $|n \rangle$. 
As a result, \autoref{eqn:secular_pop} becomes the kinetic equations governing 
the population dynamics.
For a two site model, it can be written as: 
\begin{eqnarray}
	\frac{d\tilde{\rho}_{11}(t)}{dt} &=& R_{11,11}(t) \rho_{11}(t) + R_{11,22}(t) \rho_{22}(t), \nonumber \\
	\frac{d\tilde{\rho}_{22}(t)}{dt} &=& R_{22,11}(t) \rho_{11}(t) + R_{22,22}(t) \rho_{22}(t). 
\end{eqnarray}
The transition rate from site $1$ to site $2$ is given by $k_{12}=R_{11,22}= 2 \mbox{Re}[\Gamma_{21,12}] $. Explicitly, 
\begin{eqnarray}
	k_{12} =2\kappa^2J^2\mbox{Re} \int^{t}_0 dt \, \mbox{e}^{\iu \omega_{21} t}(\mbox{e}^{ 2 \phi(t)} -1)
\end{eqnarray}
where $\omega_{21}= \omega_2 - \omega_1$. The above transition rate is the same as the prediction from FGR.


\clearpage
\newpage

\clearpage

\begin{table}[h]
\centering
\begin{tabular}{|c|c|c|c|c|c|}
\hline
    & \PME & Redfield & FGR & Experiment \\ \hline
    Rubrene & 11.1 & 41.2 &  0.33  & 3 to 15 \\ \hline
    Pentacene & 0.73 & 2.0 & 0.045 & 0.66 to 2.3 \\ \hline
    PBI-F$_2$ &  $2.2 \times 10^{-5}$ & $1.2 \times 10^{-3}$ & $2.0 \times 10^{-5}$ & - \\ \hline
    PBI-(C$_4$F$_9$)$_2$ &  0.61& 104  & 0.25  & - \\
\hline
\end{tabular}
\caption{Mobility (in cm$^2$/Vs) of organic semiconductor materials at 
   $\omega_c=500$ cm$^{-1}$, $T=300$ K and $\sigma=800$ cm$^{-1}$. } 
\label{tab:hresult}
\end{table}

\newpage

\end{document}